\documentclass[12pt,preprint]{aastex} 

\usepackage{graphicx,color}
\usepackage{amssymb,txfonts}
\usepackage{epsfig,natbib}
\usepackage{multirow}
\usepackage{rotating}
\usepackage{sidecap}
\usepackage{hyperref}
\usepackage{lscape}

\slugcomment{Draft \today}
\shorttitle{New Evidence that Magnetoconvection Drives Solar-Stellar Coronal Heating}
\begin{document}

\title{New Evidence that Magnetoconvection Drives Solar-Stellar Coronal Heating}
\author{Sanjiv K. Tiwari\altaffilmark{1,2,3,4}, Julia~K.~Thalmann\altaffilmark{5}, Navdeep K. Panesar\altaffilmark{1} Ronald L. Moore\altaffilmark{1,2}, Amy R. Winebarger\altaffilmark{1} } 
\email{sanjivtiwari80@gmail.com}
\altaffiltext{1}{NASA Marshall Space Flight Center, Mail Code ST 13, Huntsville, Alabama 35812, USA.}

\altaffiltext{2}{Center for Space and Aeronomic Research, The University of Alabama in Huntsville, 320 Sparkman Drive, Huntsville, AL 35805, USA.}

\altaffiltext{3}{Lockheed Martin Solar and Astrophysics Laboratory, 3251 Hanover
	Street, Bldg. 252, Palo Alto, CA 94304, USA. }

\altaffiltext{4}{Bay Area Environmental Research Institute, 625 2nd St. Ste 209
	Petaluma, CA 94952, USA.}

\altaffiltext{5}{Institute of Physics/ IGAM, University of Graz, Universitätsplatz 5/II, 8010 Graz, Austria.}

\begin{abstract}
How magnetic energy is injected and released in the solar corona, keeping it heated to several million degrees, remains elusive. Coronal heating generally increases with increasing magnetic field strength.  From comparison of a non-linear force-free model of the three-dimensional active-region coronal field to observed extreme-ultraviolet loops, we find that (1) umbra-to-umbra coronal loops, despite being rooted in the strongest magnetic flux, are invisible, and (2) the brightest loops have one foot in an umbra or penumbra and the other foot in another sunspot's penumbra or in unipolar or mixed-polarity plage.  The invisibility of \underline{umbra-to-umbra} loops is new evidence that magnetoconvection drives solar-stellar coronal heating: evidently the strong umbral field at \underline{both} ends quenches the magnetoconvection and hence the heating. Broadly, our results indicate that, depending on the field strength in both feet, the photospheric feet of a coronal loop on any convective star can either engender or quench coronal heating in the loop's body.
\end{abstract}

\keywords{Sun: corona --- Sun: magnetic fields --- Sun: photosphere --- Sun: transition region}

\section{INTRODUCTION}\label{intro}

Active regions have the strongest magnetic field on the Sun and contain the brightest extreme-ultraviolet (EUV) and X-ray coronal loops \citep{with77,vaia78,golu80,real14}, heated to 2-6 MK \citep{zirk93,real14} by unknown magnetic processes \citep{fish98,schr98,moor99,kats05,klim06,depo07}. As a rule, the stronger the photospheric magnetic flux the brighter the corona -- the EUV/X-ray corona in active regions is 10--100 times more luminous and 2--4 times hotter than in quiet regions and coronal holes, which are heated to only about 1.5 MK \citep{with77,vaia78,zirk93,real14} and have fields that are 10-100 times weaker than active regions \citep{wieg14}. The two most widely argued mechanisms for coronal heating are (i) magnetic-wave heating \citep[e.g.,][]{depo07,vanb11}, possibly dominant for heating the coronal plasma in quiet Sun and coronal holes, and (ii) nanoflare heating \citep{parker83a,parker88}, possibly dominant in active-region coronal loops \citep{cirt13}.  For (ii), the continuous shuffling of the field in the feet of coronal loops by the (sub-)photospheric convection supposedly braids the field in the coronal loops. This results in the corona being heated to EUV and X-ray temperatures by Ohmic dissipation of magnetic energy via reconnection in these braided loops, impulsively at many small-scale current sheets, at a rate of 10$^7$ erg cm$^{-2}$ s$^{-1}$ \citep{parker83a,parker88}.  Recent observations and coronal magnetic field modeling have provided evidence for the presence of braided loop structures and nanoflaring in the active-region corona \citep{cirt13,wine13,bros14,tiw14,thal14}.

Active regions with stronger magnetic field (often having sunspots) have brighter coronal emission than active regions with weaker magnetic field \citep{with77,vaia78,golu80,real14}. Most coronal-loop feet in sunspots are rooted in the umbra or in penumbral spines in the inner penumbra (these penumbral spines are intrusions of steeply-inclined strong umbral field into the inner penumbra \citep{tiw15aa}).  Sunspots host one foot of some of an active region's brightest coronal loops observed in EUV and X-ray wavelengths, and some of these have that foot in umbra \citep[e.g.,][]{fouk75,alis13}, consistent with sunspot umbrae having the strongest field in an active region.  Even so, umbrae often have the darkest areas of an active region in coronal EUV and X-ray images \citep{palla79,webb81,sams92,golu94}.  Here, we resolve this paradox (of why some active-region coronal loops rooted in an umbra are the brightest but others are the dimmest) by showing that the coronal EUV brightness of active-region coronal loops depends on the convective freedom in \underline{both} photospheric feet of the loops, and that considering only the loop ends rooted in an umbra and not knowing the rooting of the opposite ends, as was the case in previous studies \citep[e.g.,][]{fouk75,palla79,webb81,sams92,golu94,alis13}, resulted in this paradox.

\section{DATA AND MODELING}\label{data}

\subsection{Instrumentation and data} 
We use Solar Dynamics Observatory (SDO) Atmospheric Imaging Assembly \citep{leme12} (AIA) images, mainly from the 94 and 193 \AA\ channels, for identifying and tracking active region EUV coronal loops.  The 193 \AA\ channel of AIA detects FeXII emission at about 1.5 MK. The AIA 94 \AA\ channel is mainly sensitive to hot emission centered on an FeXVIII line (6-8 MK), but it also detects emission from about 1 MK plasma \citep{mart11,warr12,test12,boer14}.  We also inspected our active regions in the images from all other AIA channels.  

AIA UV 1600 \AA\ channel images are used for identifying locations of plage in coronal images.  The AIA 1600 \AA\ passband primarily passes lower-chromospheric continuum emission (characteristic temperature of 5000 K) but also transmits radiation from two CIV lines (at $\sim$1550 \AA) formed at 0.1 MK in the transition region. Six AIA channels are used for differential-emission-measure (DEM) analysis.  The pixel size of AIA images is 0.6\arcsec.  The time cadence for EUV AIA channels is 12s and for UV channels is 24s.  However, we have used three-minute-cadence AIA movies for the present study.  We removed solar rotation by de-rotating all the SDO images to a particular time.

Line-of-sight magnetograms from the SDO/Helioseismic and Magnetic Imager \citep{scho12} (SDO/HMI) are used to map the magnetic field strength in active regions.  HMI provides line-of-sight magnetograms at every 45s with a pixel size of 0.5\arcsec.  The cadence of the line-of-sight magnetograms used in the movies of this paper is also three-minutes.  For non-linear force-free (NLFF) field modeling, described next, SDO/HMI vector magnetograms at several selected times are used.  HMI vector magnetograms of active regions are computed using the Very-Fast-Inversion of the Stokes-Vector (VFISV) algorithm \citep{borr11hmi} and are available at a 12m cadence.

\subsection{Non-linear force-free field modeling} 
To date, there are no routine direct measurements of the coronal magnetic field.  To compensate for this lack,  NLFF magnetic field modeling based on photospheric vector magnetograms is presently the most accurate way to deduce the three-dimensional coronal magnetic field \citep{derosa09,wieg12}.  We first transform the HMI vector magnetograms to the solar disk center \citep{gary90}, i.e., deduce the vertical and horizontal vector components of magnetic field from the image-plane measurements (i.e., from the line-of-sight and transverse field).  Because the photospheric magnetic field of active regions is not completely force-free \citep{metc95,gary01,moon02,tiw12}, a preprocessing technique is applied to the HMI vector magnetograms \citep{wieg06} to achieve suitable force-free boundary conditions \citep[e.g.,][]{low85} for the coronal NLFF field modeling.  The optimization procedure is described in earlier publications \citep[e.g.,][]{wieg12optimization}. 

Since the coronal plasma is frozen to the magnetic field, the bright coronal loops trace the coronal magnetic field.  We find a great number of projected model magnetic field lines that trace the coronal loops observed in AIA 193 and 94 \AA\ images of our active regions.  This verifies that our NLFF model field sufficiently accurately approximates the true coronal magnetic field. For a coronal loop visible in the AIA images, the matching model magnetic field line gives the magnetic setting of the coronal loop's photospheric feet.  We performed modeling for each active region at every two hours during the observation period. 

It is important to note that from NLFF modeled field we need to specify only the general location of loop feet, i.e., whether a loop foot is in unipolar plage, in mixed-polarity plage, in penumbra, or in umbra.  This method is sufficiently accurate for this purpose \citep{derosa09,wieg12,wieg12optimization}.

\begin{figure}[htp]
    	\centering
    	\includegraphics[width=0.91\textwidth]{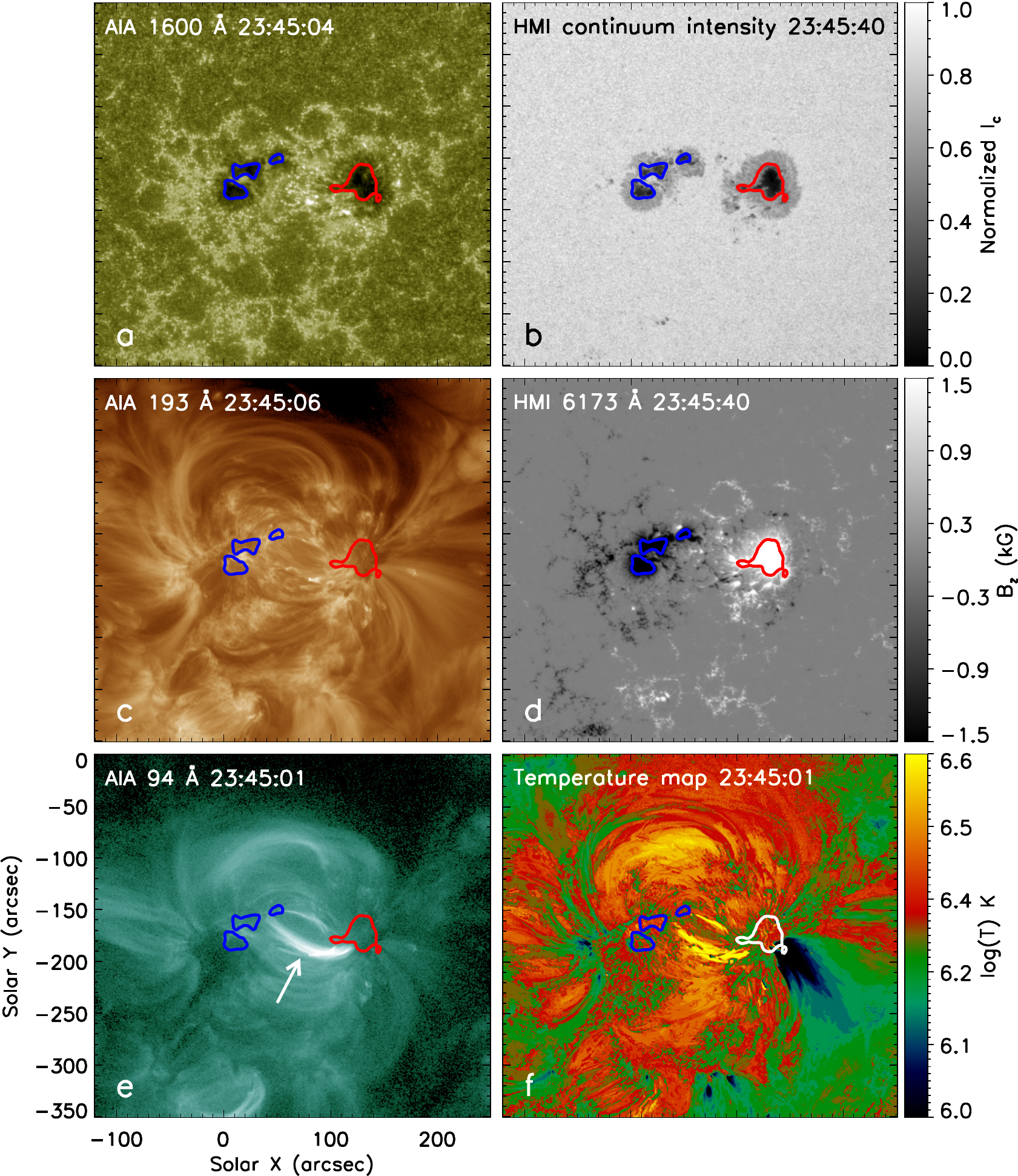}
    	\caption{\footnotesize NOAA Active Region 12108 as observed near central meridian by SDO on 07-July-2014 at 23:45UT.  {\bf a.} Photospheric and transition-region UV emission (AIA 1600 \AA).  {\bf b.} Photospheric visible-light emission (HMI continuum).  {\bf c.} Coronal EUV emission (AIA 193 \AA).  {\bf d.} Photospheric line-of-sight magnetic field (HMI 6173 \AA).  {\bf e.} Coronal EUV emission from the hottest plasma (AIA 94 \AA).  The white arrow indicates the brightest coronal loop at this time (see also Movie1, in which arrows indicate five typical examples of the brightest loops in the 94 \AA\ movie, at 02:36:01, 06:33:01, 12:36:01, 19:00:01 and 23:42:01 UT, and five typical examples of less-bright loops in the 193 \AA\ movie, at 00:09:06, 03:30:06, 13:30:06, 14:15:06 and 00:48:06 UT). All Images in the Movie1 are de-rotated to 22:00:06 UT on 01-April-2014.  {\bf f.} A temperature map of the active region obtained by DEM analysis. Blue/red (blue/white in {\bf f}) contours, which outline sunspot umbrae, are contours of $\pm$1000 G from the smoothed (by 10 pixels) line-of-sight magnetogram. }
    	\label{fig1}
\end{figure}

\section{RESULTS}\label{res}

In Fig.~\ref{fig1}, we show an example active region (NOAA 12108) in UV, EUV, continuum-intensity images and a line-of-sight magnetogram.  This active region has a simple bipolar field configuration and contains sunspots of opposite magnetic polarity. The active region was flare-quiet -- it produced no C-class or larger flares within 24 hours before and after our observations. We selected a non-flaring active region because we are interested in quasi-steady heating and not in the large bursts of heating by flares. During the 26 hours of our observations, the active region passed the central meridian 8$^\circ$ south of disk center, and so was always close to disk center, thus avoiding projection errors.  Coronal loops of different temperatures can be identified in the coronal EUV images of SDO/AIA.  Noticeably, the sunspots of the active region appear dark in continuum intensity (Fig.~\ref{fig1}b) and UV images (Fig.~\ref{fig1}a).  The sunspots are surrounded by weaker-field plage regions, which are bright in the AIA 1600 \AA\ image (Fig.~\ref{fig1}a).  The loops are rooted in different photospheric magnetic features such as sunspots, unipolar plage, and mixed-polarity plage (compare Fig.~\ref{fig1}c, d, and e).  The temperature map (Fig.~\ref{fig1}f), derived from taking into account the emission across the temperature regime $\sim$0.6 to 6 MK (corresponding to AIA 171 to 94 \AA\ channels) through DEM, following \cite{asch13}, shows the approximate temperatures of coronal loops.  The temperature map shows that the brightest coronal loops in the 94 \AA\ image are the hottest ones. We emphasize that, although the DEM analysis indicates that the brightest AIA 94 \AA\ loops are the hottest, our main results of the paper do not depend on the DEM calculations.

\begin{figure}[h]
	\centering
	\includegraphics[width=1.015\textwidth]{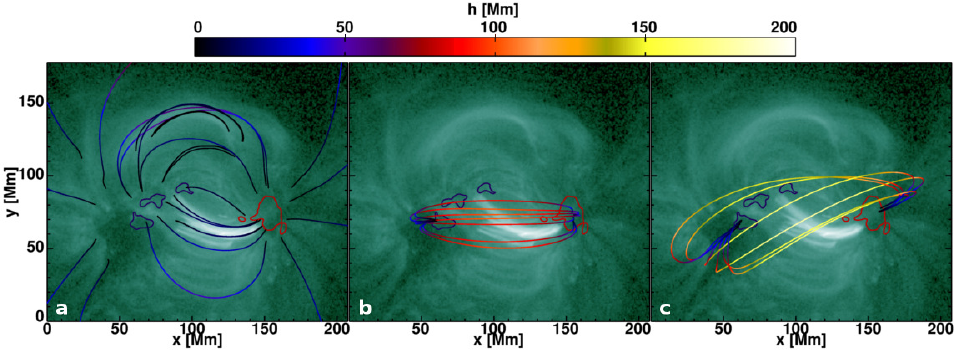}
	\caption{\footnotesize Model coronal magnetic field of NOAA Active Region 12108 over plotted on AIA 94 \AA\ image.  The NLFF field model is based on HMI photospheric vector magnetogram recorded at 23:36:00 UT.  The background AIA 94 \AA\ image is the nearest in time, at 23:36:01 UT.  The blue/red contours here and in Movie1 are made the same way as in Fig.~\ref{fig1}.  The height at each point along each model field line is color-coded.  {\bf a.} Sample model magnetic field lines matching the coronal loops in 94 \AA\ and/or 193 \AA\ images.  {\bf b.} A set of lower-altitude model umbra-to-umbra field lines having apex heights of 90—-100 Mm.  {\bf c.} A set of higher-altitude model umbra-to-umbra field lines having heights of 150—-175 Mm.  The higher/longer umbral field lines are naturally rooted near outer edge of each umbra, in or close to penumbral spines in the inner penumbra.  None of the umbra-to-umbra model field lines match any discernible coronal loops.  See Movie2 for a three-dimensional view of these umbra-to-umbra model field lines.}
	\label{fig2}
\end{figure}

\begin{figure}[htp]
	\centering
	\includegraphics[width=0.91\textwidth]{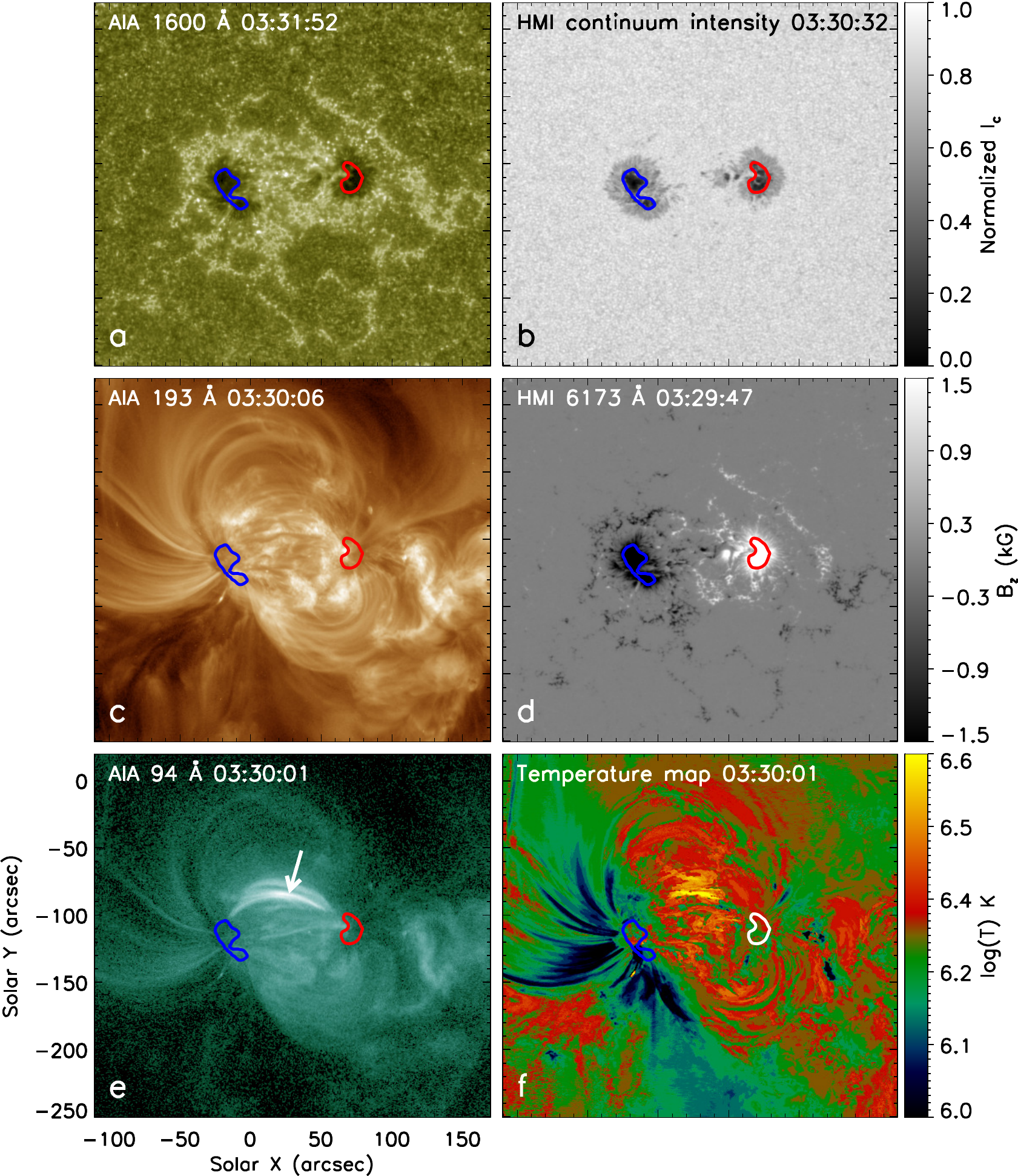}
	\caption{\footnotesize Similar figure as to Fig.~\ref{fig1} but for second example NOAA Active Region 12021 observed on 01/02-April-2014, when it was about as near disk center as the first active region was during its observations.  Contours have the same levels and meaning as in Fig.~\ref{fig1}.  An arrow in {\it e} again points to the brightest and hottest loop at this time (see also Movie3, in which arrows indicate five typical examples of the brightest loops in the 94 \AA\ movie, at 12:30:01, 13:18:01, 20:57:01, 21:21:01 and 04:39:01 UT, and five typical examples of less-bright loops in the 193 \AA\ movie, at 10:00:06, 01:12:06, 02:30:06, 06:21:06 and 09:54:06 UT). All images in the Movie3 are de-rotated to 11:30:40 UT on 07-July-2014.}
	\label{fig3}
\end{figure}

\begin{figure}[h]
	\centering
	\includegraphics[width=1.015\textwidth]{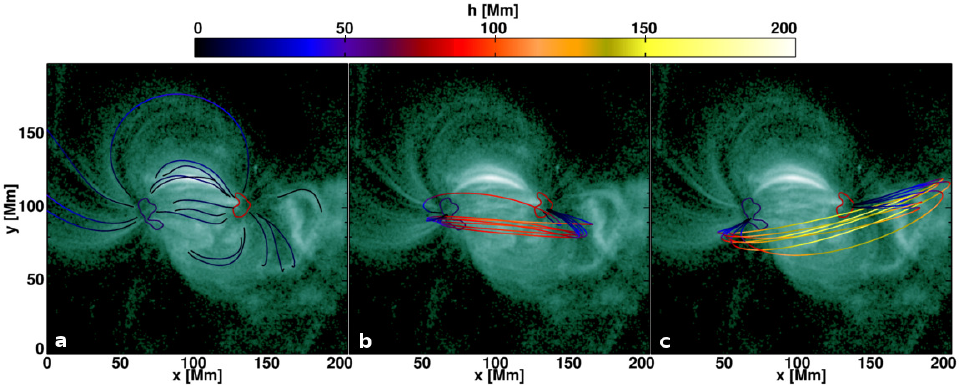}
	\caption{\footnotesize Model active-region coronal magnetic field of NOAA Active Region 12021. The NLFF model field is based on the HMI vector data observed on 02-April-2014 at 03:35:25UT, and the AIA 94 \AA\ background image is the nearest in time at 03:36:01UT\@.  See Movie4 for a three-dimensional view of these umbra-to-umbra model field lines.}
	\label{fig4}
\end{figure}

We use sequences of AIA 193 and 94 \AA\ images, such as those shown in Fig.~\ref{fig1}, to study the evolution of bright active-region coronal loops (Movie1).  In these sequences:  (i) Bright EUV coronal loops are episodic, with lifetimes of 1-2 hours.  (ii) Many of the brightest EUV loops are apparently rooted within or at the edge of a sunspot umbra at one foot, with their other foot located in opposite-polarity plage or penumbra.  (iii) Some of the other brightest loops have one foot rooted in umbra or penumbra and the other foot in mixed-polarity flux.  (iv) The bright coronal loops connecting opposite unipolar plage regions are never as bright as those with one foot rooted in a sunspot or in mixed-polarity plage.  (v) Coronal loops with each foot in a sunspot umbra of opposite polarity (``umbra-to-umbra" loops) are \emph{never} visible, neither in the wavelength bands presented here, nor in any of the other AIA passbands.

To examine whether umbra-to-umbra coronal magnetic loops are present, we use a NLFF field model of the active-region coronal magnetic field, based on a SDO/HMI photospheric vector magnetogram.  In Fig.~\ref{fig2}a, selected model field lines are shown overlaid on the nearest-in-time AIA 94 \AA\ image.  Each of these model field lines traces an observed loop closely, validating the model field.  Despite the absence of any discernible connection between the two sunspot umbrae in EUV coronal images, the three-dimensional model magnetic field shows that there are field lines connecting the opposite-polarity umbrae of the active region (Movie2).  In Figs.~\ref{fig2}b, c, we plot two sets of the model field lines that link the two opposite-polarity umbrae.  Since the length of a coronal loop is an important factor in determining the loop's coronal brightness -- shorter loops requiring lesser amount of heating than longer ones for the same brightness \citep{mand00,klim06,real14,kano14} -- we display umbra-to-umbra loops of different apex heights (and therefore of different lengths).  The two sets of umbra-to-umbra loops displayed in Figs.~\ref{fig2}b and \ref{fig2}c have apex heights $\sim$90-100 mega-meters (hereafter Mm), ``low-apex" fields, and $\sim$150-175 Mm, ``high-apex" fields, respectively.  Neither low-apex (short) nor high-apex (long) loops that connect the sunspot umbrae are visible in the AIA 94 and 193 \AA\ images. We also verified the invisibility of these loops in the movies from all other AIA channels.  This indicates that the coronal loops connecting opposite-polarity umbrae have the lowest coronal-temperature plasma density (and thus the lowest coronal brightness), and hence are the least heated \citep{vaia78}. It is worth mentioning here that a dim umbra-to-umbra loop seen in AIA 171 \AA\ images of the active region (NOAA 12108) two days after our observations \citep{chit16}, is reasonable because a light bridge, which is a signature of magnetoconvection in the umbra, was present then in the leading-polarity sunspot, and was at the foot of that loop.

In Fig.~\ref{fig3}, we show images and the temperature map of another flare-quiet active region (NOAA 12021), the model field lines of which are shown in Fig.~\ref{fig4}.  We find the same pattern for the EUV brightness of coronal loops with respect to their magnetic rooting within the active region.  Again, the modeled umbra-to-umbra field lines do not match any discernible EUV coronal loops within an extended 24-hour time span (Movie3). Although here we show $\sim$24-26 hours of data for only two examples, we have used JHelioviewer Software to find many similar active regions having a large sunspot of each polarity with no light bridges in the umbrae. These active regions showed no exception to our results presented here, e.g., we do not find any visible umbra-to-umbra coronal loops.

\section{Discussion and Conclusions}


We interpret our observational and modeling results as follows.  The convective freedom at \underline{both} photospheric feet is the primary determinant of the brightness of active region coronal loops.  Umbra-to-umbra loops are the dimmest (invisible) because the convection in each foot of these loops is so strongly suppressed by the strong umbral fields that the magnetic energy input to those loops is quenched.  That is, field braiding for powering heating by magnetic reconnection is quenched in these coronal loops. On the one hand, the EUV invisibility of umbra-to-umbra loops and the observed magnetic rooting of the bright EUV loops are consistent with Parker's idea \citep{parker83a} that convection shuffles field lines and the subsequent small-scale reconnection \citep[nanoflares:][]{parker88} in the resultant braided coronal loops \citep{cirt13,wine13,bros14,tiw14,thal14,pont17} releases magnetic energy to heat the corona, which implies that the strong suppression of convection by the strong field in umbrae should quench coronal heating in umbra-to-umbra loops. On the other hand, any appreciable heating from magnetic waves generated by convection-churning of the loop feet is evidently also suppressed in umbra-to-umbra coronal loops.

\begin{figure}[tp]
	\centering
	\includegraphics[trim=2.5cm 4.5cm 2.5cm 1.5cm, clip,width=0.92\textwidth]{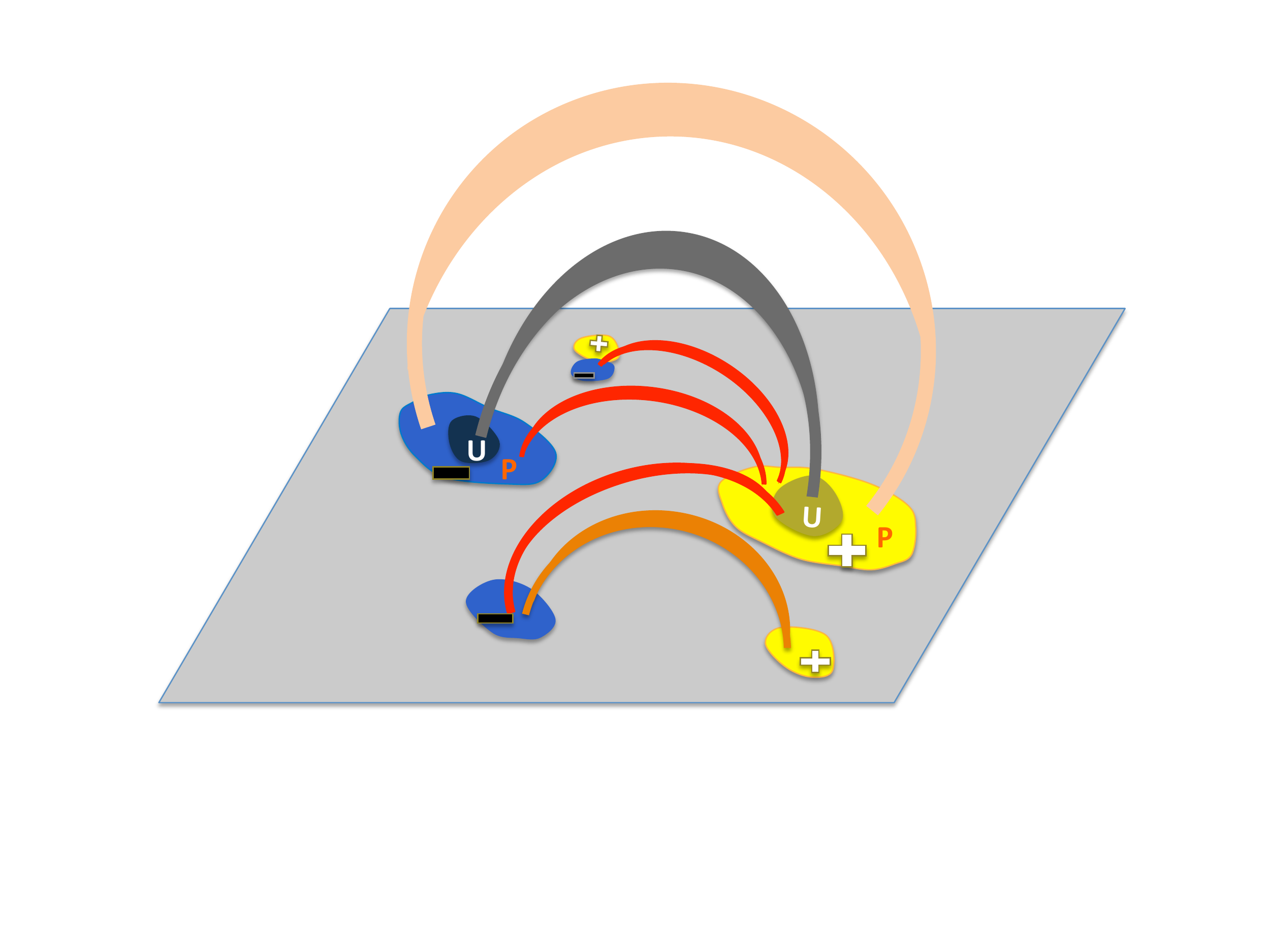}
	\caption{\footnotesize A schematic drawing depicting the dependence of the coronal EUV brightness of active-region coronal loops on the photospheric magnetic rooting of the loops.  Yellow/blue colors show positive/negative polarity of surface magnetic flux. `U' and `P' stand for sunspot umbra and penumbra, respectively. Different colors of different loops indicate their brightness, with red/dark-gray being the brightest/dimmest, and with bright orange being dimmer than red and brighter than the pale orange.  Each of the two foreground positive and negative magnetic areas is a unipolar plage region.  A mixed polarity plage region is present on the left in the background, where one of the brightest loops has one foot, the other foot being in sunspot penumbra of positive polarity.  The taller of the two penumbra-to-penumbra coronal loops is the dimmer, presumably because it is longer.}
	\label{fig5}
\end{figure}

Because the brightest coronal loops are the ones linking an umbra or inner penumbra to unipolar plage, mixed-polarity plage, or penumbra, we infer that they are the brightest because they have one foot rooted in umbra or inner penumbral spines and therefore have stronger magnetic field in the loop body than in plage-to-plage loops of the same length (which never become as bright as the brightest loops).  Plausibly, because of that and because the other foot has vigorous convection, these loops with only one end in umbra have a higher rate of magnetic energy injection (higher Poynting flux) and higher dissipation rate via reconnection, and thus brighter coronal emission than the plage-to-plage loops.  All previous studies of the brightness of coronal loops stemming from an umbra considered only the loop feet in that umbra and did not know the magnetic settings of the other ends of the loops. Thus, they did not discern why some active region coronal loops rooted in umbra are the brightest \citep{fouk75,alis13} and others are the dimmest \citep{webb81,palla79,sams92,golu94}.  The sketch in Fig.~\ref{fig5} depicts what we infer to be the dependence of the brightness of coronal loops on their magnetic rooting in an active region.  \emph{Our observations imply that so long as the field can be braided by convection in a loop foot, the stronger the field in the loop the stronger the coronal heating.} Our results qualitatively support the models of \cite{hurl02}, and \cite{chen14}.   

Thus, magnetoconvective driving at \underline{both} feet of a loop is the key. Whether the subsequent transport and dissipation mechanism is braiding/nanoflare or wave heating is beyond this investigation. Irrespective of whether  active-region corona is heated more by dissipation of magnetic waves or more by nanoflares, our results indicate that, depending on the field strength in \underline{both} feet, the photospheric feet of any coronal loop on the Sun -- or on any star with a convective envelope -- can either engender or quench coronal heating in the loop's body.  Our results are a new line of evidence for the widely held view that the quasi-steady coronal heating in active regions is driven by magnetoconvection in the feet of the loops \citep{parker83a,parker88}.  The alternative possibility, that the heating mostly comes from dissipation of free magnetic energy already in the loop field when it emerged from inside the Sun \citep{fish98}, is unlikely in view of the EUV invisibility of umbra-to-umbra loops.

We observed, as have a few other researchers \citep{falc97,chit17}, that the active-region coronal loops that are most persistently the brightest are rooted in regions of mixed-polarity magnetic flux at one end or at both ends. This indicates that the evolutionary interaction of the opposite-polarity flux, which is driven by magnetoconvection in and below the photosphere, somehow enhances the rate of injection of free magnetic energy into the accessed coronal loops, in excess of what the rate would be were the magnetic flux unipolar at both ends.


\acknowledgments
We thank Allen Gary, Alphonse Sterling, David McKenzie, Davina Innes and P. Venkatakrishnan for useful comments on the manuscript. S.K.T., R.L.M. and A.R.W. were supported by funding from the LWS-TRT program and the HGI Program of the Heliophysics Division of NASA's SMD. N.K.P.'s research was supported by an appointment to the NASA Postdoctoral Program at the NASA/MSFC, administered by USRA under a contract with NASA. JKT acknowledges support from Austrian Science Fund (FWF): P25383-N27. We acknowledge using NASA/SDO-AIA-HMI data and the JHelioviewer Software.


\end{document}